\documentclass[11pt]{iopart} % IOP journal style

\usepackage{graphicx}    % Include figure files
\usepackage{latexsym}    % 
\usepackage{color}       % 
\usepackage{dcolumn}     % Align table columns on decimal point
\usepackage{bm}          % bold math
\usepackage{amssymb}     % for math
\usepackage{hyperref}    % add hypertext capabilities

\usepackage{graphics}
\usepackage{epsfig}
\usepackage{caption}
\usepackage{subfigure}
\usepackage{nicefrac}

\begin{document}

\title[Numerical study of impeller-driven von K\'arm\'an flows]
{Numerical study of impeller-driven von K\'arm\'an flows via a volume penalization method}

\author{S Kreuzahler$^1$, D Schulz$^1$, H Homann$^2$, Y Ponty$^2$ and R Grauer$^1$}
\address{$^1$ Institut f\"ur Theoretische Physik I, Ruhr-Universit\"at Bochum, 44780 Bochum, Germany}
\address{$^2$ Universit\'e de Nice-Sophia, CNRS, Observatoire de la C\^ote d'Azur, 
B. P. 4229 06304 Nice Cedex 4 , France}

\begin{abstract}
The von K\'arm\'an flow apparatus produces a highly turbulent flow inside a cylinder vessel 
driven by two counter-rotating impellers. 
Over more than two decades, this experiment has become a very classic turbulence tool, 
studied for a wide range of physical systems by many groups, 
with incompressible flow,  compressible flow, for magnetohydrodynamics and dynamo studies 
inside liquid metal, for particle tracking purposes, and recently with turbulent super-fluid helium.
We present a direct numerical simulation (DNS) version the von K\'arm\'an flow, 
forced by two rotating impellers.  
The cylinder geometry and the rotating objects are modelled via a penalization method 
and implemented in a massive parallel pseudo-spectral Navier-Stokes solver. 
We choose a special configuration (TM28) of the impellers 
to be able to compare with set of water experiments well documented. 
But our good comparison results implied, that our numerical modelling could also be applied 
to many physical systems and configurations driven by the von K\'arm\'an flow.  
The decomposition into poloidal, toroidal components 
and the mean velocity fields from our simulations are in agreement with experimental results. 
We analyzed also the flow structure close to the impeller blades 
and found different vortex topologies. 

\end{abstract}


% 47.27.ek= DNS;   
% 47.11.Kb, 47.27.er = spectral methods; 
% 47.27.E- = turbulence simulation and modeling
% 47.65.-d 	Magnetohydrodynamics and electrohydrodynamics
% 47.11.-j = computational methods;  ? not found in base IAPS  (old pacs number ?) 

%\begin{keyword}
%% keywords here, in the form: keyword \sep keyword
\noindent{\it Volume penalization \/ Pseudospectral method \/ Moving boundaries \/ von K\'arm\'an flow}
%% MSC codes here, in the form: \MSC code \sep code
%% or \MSC[2008] code \sep code (2000 is the default)

%\end{keyword}

\maketitle

\section{Introduction}\label{sec:intro}

A von K\'arm\'an experiment is a cylindrical vessel in which a flow is generated by the
rotation of two impellers at the extremities of the vessel \cite{vonkarman}.
Studying turbulence by means of von K\'arm\'an fluid experiments has a strong
tradition in last two decade. Several teams set up such experiments with different
designs in incompressible \cite{dijkstra_flow_1983,douady_direct_1991}
and compressible flows
\cite{fauve_pressure_1993,pinton_1994,abry_analysis_1994,labbe_study_1996}.
This type of experiments was also one of the first setups used to
study Lagrangian statistics of turbulent flows
\cite{mordant_measurement_2001,mordant_long_2002,la_porta_using_2000,la_porta_fluid_2001},
by tracking solid particles or bubbles. Recently, a helium super-fluid 
experiment has been built with the classic von K\'arm\'an configuration \cite{shrek1}  
to reach even higher Reynold number, and to study the interaction of the super and classic fluid. 

In the last decade, in order to gain a better understanding of the
underlying processes of magnetohydrodynamic and dynamo effect,  many experimental groups have investigated 
experiments using liquid sodium~\cite{gailitis2000, gailitis2001, muller2000, stieglitz2001}.
A very successful experiment used the von K\'arm\'an apparatus with Sodium liquid (VKS) hosted
in Cadarache which was able to reproduce dynamo action in a turbulent
flow \cite{monchaux2007,berhanu2007,monchaux_von_2009,berhanu_bistability_2009,
  berhanu_dynamo_2010,gallet_experimental_2012}. 
Before starting with sodium experiments, prototypes filled with water were
set up, which are smaller than the final VKS machine by a factor of
two.  They compared and optimized different impeller designs to seek
the highest kinematic dynamo growth rate
\cite{marie_numerical_2003,ravelet_toward_2005}.

In this paper we will focus on the purely hydrodynamic properties of
the von K\'arm\'an flow driven by rotating impellers and keep the
investigation of the magnetized dynamics, or other experimental application as a future goal.

In this work we perform direct numerical simulations of a VKS flow,
thus a three-dimensional impeller-driven turbulent flow in cylindrical
geometry. The geometry of rotating impellers assembled of several
basic geometric objects is modeled via an immersed boundary technique
(IBM) and implemented in a massive parallel pseudo-spectral
Navier-Stokes solver. High resolution simulations allow for the
development of a fully developed turbulent flow.  We first compare our
numerical results with the water campaigns of the Saclay group
\cite{marie_these_2003, marie2004, ravelet_multistability_2004,
  ravelet_these_2005, ravelet_supercritical_2008} to validate our
numerical modelization approach. We will also study the flow dynamics
in the vicinity of impellers, a region not easily accessible
experimentally.

\section{Numerical method}\label{sec:sim}

\subsection{Basic equations}
We consider the incompressible Navier-Stokes equation

\begin{eqnarray}
\label{eq:ns}
\frac{\partial \mathbf{u}}{\partial t} + (\mathbf{u} \cdot \nabla) \mathbf{u} = - \nabla p + \nu \Delta \mathbf{u}
\end{eqnarray}
with the velocity field $\mathbf{u}(\mathbf{x}, t)$, pressure $p$,
viscosity $\nu$.  The velocity field additionally fulfills the
incompressibility condition $\nabla \cdot \mathbf{u} = 0$.  At the
boundaries we impose a no-slip condition such that the velocity of the
fluid equals the velocity of the boundary itself
$\mathbf{u}|_{boundary} = \mathbf{V}_{penalized}$. It is zero on the
fixed outer cylinder and equals the solid rotation velocity on the
disks and the blade structures. The cylindrical wall and the moving
impellers are implemented by a penalization method (see
\ref{sec:penalty}).

\subsection{The Fourier-spectral scheme}
To solve the equation system a standard pseudo-spectral method is
applied using the $\nicefrac{2}{3}$ rule for dealiasing and
resolutions of $256^3$ and $512^3$ grid points. The time derivative on
the left-hand side of the Navier-Stokes equation is discretized via a strongly stable
third order Runge-Kutta method \cite{shu1988}.  Incompressible turbulent flows have
been intensively studied in a periodic box, a classical mathematical
framework for theories \cite{book_uriel} as well as for numerical
simulations of isotropic and homogeneous turbulence
\cite{patterson,vincent1991}.  In this geometry the pseudo-spectral
numerical method is the most precise global numerical method for a
fixed mesh size and the success of this method is essentially due to the
efficiency of the Fast Fourier Transform.  In the present work, we are
using an immersed boundary method to impose no-slip boundary
conditions inside the simulation domain.  In this framework we lose
the spectral precision near the boundaries, but we are able to design
any geometry of static or moving structure while keeping the usability
of a pseudo-spectral code.  The implementation of the penalization
boundaries and the moving impellers increase the CPU time by $2$
times. The most time-consuming part of this implementation is the recalculation
of the moving boundaries in each step. For future MHD simulations
we expect a factor below $2$, as the penalization then
comsumes less time compared to the solution of the basic
equations.  Those methods are sufficiently accurate to reproduce standard
benchmarks \cite{minguez_high-order_2008} and benchmarks depending
crucially on the boundary layer solution
\cite{homann_effect_2013}. The used method will now be explained in
more detail.

\subsection{Penalization method}\label{sec:penalty}

To simulate flows within a solid cylindrical boundary and moving
impellers with complex shape, a penalization method is applied. For
points inside the boundaries a ``pseudo'' forcing term is added to the
right-hand side of the Navier-Stokes equation~(\ref{eq:ns}), which
adjusts the velocity exactly to the prescribed value in and on the
wall or the impellers.  The method we used to calculate the force,
which is first introduced in \cite{mohd1997,fad2000}, is called direct
forcing and allows to calculate the force directly from the
Navier-Stokes equation without the necessity of further auxiliary
parameters. To increase the precision of the boundary layers, we used
a predictor for the pressure gradient \cite{brown_accurate_2001}.

As we deal with complex geometries the boundaries of the objects do
not coincide with the rectangular grid. This makes it necessary to
interpolate at the boundary, for which we take the volume fraction
$V_b$ occupied by the solid object into account. Regarding the volume
of each cell $V_c = n^3 \Delta x \Delta y \Delta z$, the force is
weighted with the factor $\nicefrac{V_b}{V_c}$. Practically, this is
performed via a Monte-Carlo method using 50 random points within each
cell to calculate the volume \cite{fad2000}.  The details of this
penalization method were described and tested in another fluid context
\cite{homann_effect_2013}.

The solid rotating velocity of the impellers is simply given by the
angular velocity $\mathbf{\Omega}$ and the distance from the axis
$\mathbf{r}$ as $\mathbf{V}_{boundary} = \mathbf{\Omega} \times
\mathbf{r}$. The angular velocity of the impellers can be set
independently for each impeller.

\subsection{Configurations of our numerical experiments}\label{sec:conf}

\subsubsection{Numerical experiment setup}
\label{sec:setup}
Our aim is to model a configuration similar to the von K\'arm\'an
experiments in which a cylindrical vessel is filled with liquid.  The
fluid is driven by two counter-rotating impellers, one on each side of
the vessel. Each impeller consists of a disk on which several blades
are mounted.

We create an embedded cylinder inside the periodic box, using our
penalization method, with a radius $R_c=3.0$. Though the periodicity is kept
along the z axis to decrease Gibbs oscillations, the velocity at the end of
the cylinder is close to zero due to the symmetry of the system.  The height of
the cylinder is $2\pi$, giving an aspect ratio of $2\pi/R_c$.  In
experimental setups this height varies from 2 up to 3 cylinder radii.
The interior of this cylinder represents $\sim 71 \%$ of the total
computing domain.
     
For one set of simulations we choose a very similar configuration of
the curved disk-blades to the setup called ``TM28''
\cite{marie_numerical_2003} with a distance between the disks of $1.8
R_c$, a radius of the rotating disk of $R_d=0.9 R_c$, a height of the
eight blades of $0.2 R_c$ and a curvature radius of the blades of
$C=0.5 R_c$.  The angle of the expelled flow at the end of the blades
is given by $ \alpha = \mbox{arcsin} \left ( \nicefrac{R_d}{2 C}
\right ) \approx 1.11976~rad \sim 64.15 ~deg$.  The simulated straight
blade configuration is close to the ``TM70'' and ``TM80''
configuration \cite{marie2004,ravelet_these_2005,ravelet_toward_2005},
with eight blades per disk. A difference is that our disk radius is
$R_d=0.9 R_c$ instead of ($R_{TM70}=0.75 R_c$, $R_{TM80}=0.95
R_c$). We call this configuration {straight} configuration. Here,
$C=\pm \infty$ and $\alpha=0$.

In the following we analyze these two blade geometries (TM28 curved
and straight blades). The disks are always counter-rotating with the
same rotation rate. Depending on the rotation direction we have two
different blade curvatures (called $+$ or $-$) (see figure
\ref{fig:setupcurved}).

\includegraphics[width=12cm]{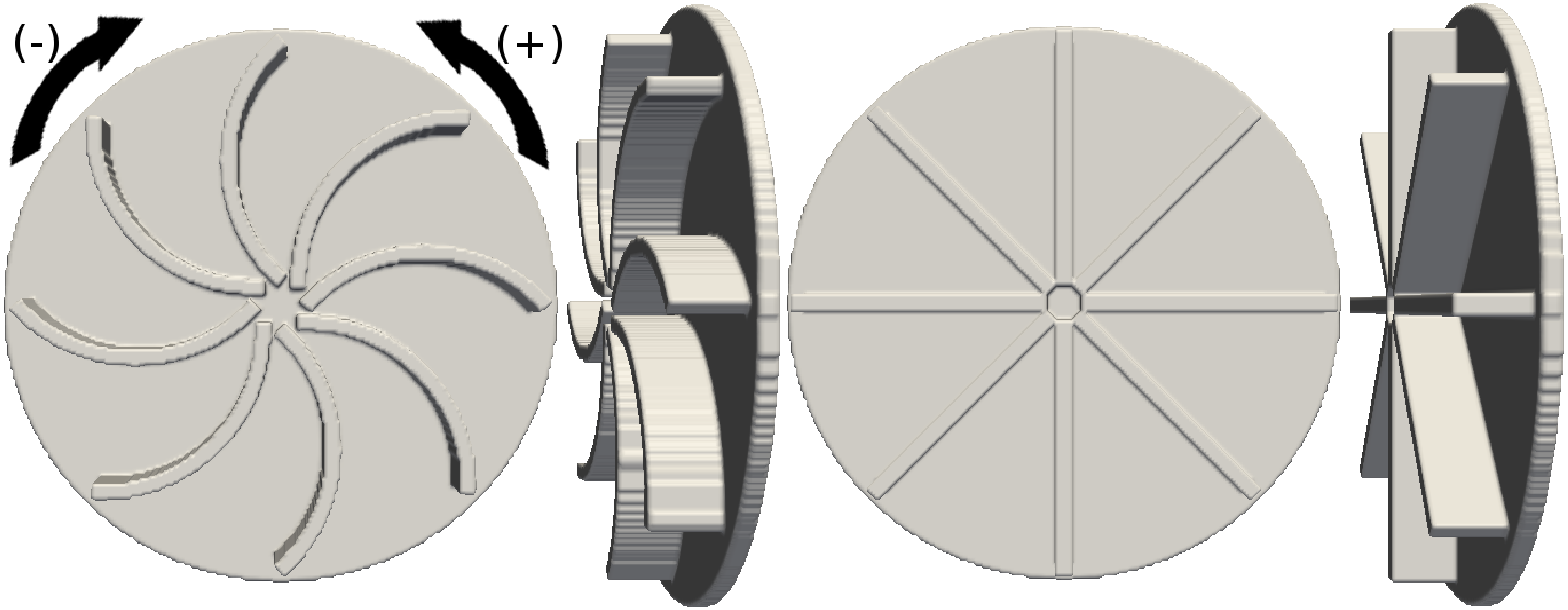}
\captionof{figure}[Simulated impellers]{Simulated impellers with 8
  curved blades (left) and 8 straight blades (right): A positive
  curvature, denoted (+), means that the convex side of the blade
  points in the turning direction of the impeller, while for a
  negative value, denoted (-), it is the concave side. }
\label{fig:setupcurved}

\subsubsection{Non dimensional numbers}

We have chosen six simulations to highlight our results. We vary the
viscosity $\nu$ controlling the dissipative term, the rotation rate
$\Omega$ of the impellers, and the curvature of the blades. Of course
other parameters such as the geometry of the blade, the number of
blades, the difference in the rotation rate of the disks might change
the topology of the flow and the quantitative results.  %Our
%configurations of the impellers are fully described above
%\ref{sec:setup}.

From our numerical data we compute a set of meaningful non-dimensional
numbers (see table \ref{tab:datas}). The Reynolds number accessible
with direct numerical simulations with $512^3$ grid points is
as usual a few orders of magnitude lower than experimental
Reynolds numbers. Nevertheless, at this Reynolds number the flow is
turbulent. Our measured fluctuation level $ \delta =
\nicefrac{\overline{\langle \mathbf{u}^2 \rangle}}{\langle
  \overline{\mathbf{u}}^2 \rangle}$ defined in
\cite{cortet_normalized_2009} is similar to that obtained from the water
experiments for synchronized rotating disks and with an annulus
deviator configuration in the ``TM60'' and ``TM73'' setup.  With
asynchronous disk rotations or without annulus, this level of
fluctuation could be higher (above $2.0$)
\cite{cortet_normalized_2009}.

The efficiency $E_f= \nicefrac{U_{max}}{\Omega R_d}$ of the impellers,
which measures how much energy is injected into the fluid, is defined
as the maximum velocity of the fluid in the bulk induced by the
impeller motion.  The variation of the $E_f$ as a function of the
expelled flow angle $\alpha$ for different experimental configurations
is confronted with our numerics in figure \ref{fig:gamma-Ef}.b.  With
a Reynolds number three orders of magnitude lower than the experiments, we
nevertheless have a good agreement with the experimental
efficiency. The efficiency number decreases with the same slope found
in the water experiments.  We noticed that our straight blade
efficiency is almost identical with the ``TM70'' configuration
efficiency.  For the positive curved blades our numerical data is
closer to the ``TM60'' configuration (16 blades) than to the ``TM28''
(8 blades).

The ratio of the root mean square velocity and the maximum velocity of
the impellers ($U_{rms}/V_{max}$) is pretty close to the TM28
experiment for our higher rotation rate simulation
($1.6\Omega=2.4$). 

Another non-dimensional number is the ratio of rotation period of the
impellers $T_{\Omega}=1/f=\frac{2 \pi}{\Omega}$ and the large eddy
turn over time $T_{nl}$. Our simulations are in the same disk rotation
regime as the experiments (see the last line of table
\ref{tab:datas}) showing that there is a bit more than two large eddy
turn over times during one turn of the impellers.

\begin{table}
\begin{tabular}{|c|c||c|c|c|c|c|c| }
\hline
    & TM28              & curved (+) & curved (-)  & \tiny{(straight)} & (+) 512  & (+) $1.6\Omega$ &  (+) $\nu/2$   \\
\hline
\mbox{grid size} & ---      &  $256^3$   &  $256^3$    & $256^3$   &  $512^3$ &  $512^3$        &  $512^3$   \\
\hline
$\nu$    &  $10^{-6} ~m^2/s$ &  $0.005$   &  $0.005$   &  $0.005$  &  $0.005$  &  $0.005$       &  $0.0025$  \\
\hline
$\Omega$ &  $28.4628 s^{-1}$ & $1.5$      &   $1.5$    &  $1.5$    &   $1.5$   &   $2.4$        &    $1.5$   \\ 
\hline
$U_{rms}$ &  $1~ m/s$        &   $0.9397$ &  $1.0641$  &  $1.0843$ &  $0.9429$ &  $1.5352$      &  $0.9619$  \\
\hline
$U_{max}$ &                  &  $2.268$   &  $3.143$   &  $2.859$  &  $2.268$  &  $3.616$       &   $2.268$  \\
\hline
$L$      & $0.1$ m          &  $1.6831$  &  $1.9011$  &  $1.9149$ &  $1.6349$ &  $1.9277$      &  $1.9038$  \\
\hline
\hline
$T_{nl}$  &    $0.1 $ s      & $1.7911$   &  $1.7865$  &  $1.7660$ & $1.7339$  &  $1.2556$      &  $1.9792$  \\
\hline 
$R_{exp}$ &   $2.84~10^5$    &  $2430$    &   $2430$   &  $2430$   & $2430$    & $3888$         & $4860$     \\
\hline
$R_{num}$ &  $ 10^5 $        &  $316$     &   $404$    &   $415$   & $308$     &  $591$         &  $732$     \\
\hline
$E_f$   & $0.64$            & $0.50$    &  $0.698$    & $0.635$   & $0.504$   &  $0.502$       & $0.504$    \\
\hline
$\delta$ & $1.5-2.2$        & $1.452$   & $1.433$    & $1.487$    & $1.40$    & $1.519$        & $1.519$    \\
\hline
$U_{rms}/V_{max}$ &  $0.40$   &  $0.232$   &   $0.262$ &  $0.267$   & $0.232$   &   $0.379$      & $0.237$    \\
\hline
$\nicefrac{T_{\Omega}}{T_{nl}}$
                 &  $2.207$   & $2.34$   &  $2.338$  &  $2.3446$  &  $2.371$  &  $2.0850$     &  $2.1164$  \\
\hline
\end{tabular}
\caption{
Simulation quantities are compared with water experiments results 
from \cite{marie_numerical_2003,marie2004,cortet_normalized_2009} 
and specially the configuration ``TM28''. 
We collected and defined several quantities or non-dimensional numbers: 
$\nu$ the kinematic viscosity of water or in our simulations, 
$\Omega$ the rotation rate, 
$U_{rms} = \overline{\sqrt{2E(t)} }$, 
$U_{max}$ is the (spatial) maximum of the (time-averaged) mean velocity field in the bulk 
in the range $-0.8R_c<z<0.8R_c$,  where z=0 is the center of the cylinder.
$L=\frac{2 \pi}{\sum E(k)} \sum E(k)/k$ is the integral length scale computed with $E(K)$, 
the isotropic spectral density of the kinetic energy, 
and $T_{nl} = L/U_{rms}$ is the eddy turn over time.
We used the experimental Reynolds number definition of VKE-VKS experiments
$R_{exp}  = \nicefrac{\Omega R_d R_c}{\nu}$.
To compare with numerical works, we define another Reynolds number $R_{num}=\nicefrac{U_{rms} L }{\nu}$. 
The efficiency $E_f= \nicefrac{U_{max}}{\Omega R_c}$ of the impellers states how much energy is injected into the fluid.
$ \delta = \nicefrac{\overline{\langle \mathbf{u}^2 \rangle}}{\langle \overline{\mathbf{u}}^2 \rangle}$ 
is the fluctuation level defined in \cite{cortet_normalized_2009}. We present also the ratios $U_{rms}/V_{max}$
and $T_{\Omega}/T_{nl}=\frac{2 \pi U_{rms}}{\Omega L }$. 
Our simulations are during more than $20 T_{\Omega}$ (turns of the impellers), 
which is the duration of the time average computation.   
}

\label{tab:datas}
\end{table}

\section{Mean bulk flow structure}\label{sec:global}

\subsection{Global flow profiles}\label{sec:profiles}
\begin{figure}[hb!]
\includegraphics[width=18cm]{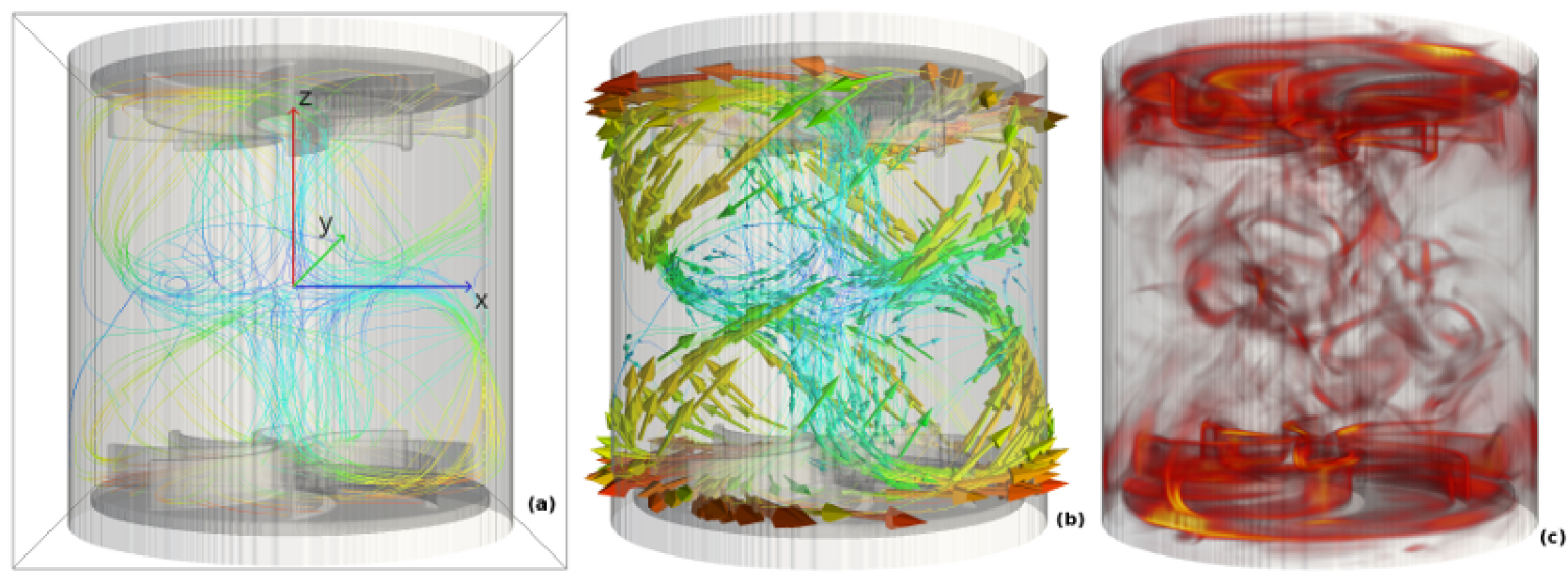}
\caption{(a) Streamlines and (b) vector field of the time-averaged
  flow and (c) a snapshot of the vorticity with (+) configuration.}
\label{fig:visu3d}
\end{figure}

While integrating the Navier-Stokes equations we additionally time
averaged the velocity field. The stream lines and the vector field of
the mean flow $\overline{\mathbf{u}}$ are shown in
figure \ref{fig:visu3d}.a and figure \ref{fig:visu3d}.b, respectively.
These images reproduce the classical images of a S2T2 flow of the von
K\'arm\'an flow.  For comparison, a snapshot of the enstrophy
(figure \ref{fig:visu3d}.c) shows interacting vortex filaments in the
bulk region which is characteristic to a turbulent flow.  The
vorticity is produced and thus very high near the blades. This observation will be
analyzed in detail in section \ref{sec:local}.

\subsection{Poloidal and toroidal components}\label{sec:poltor}
For further analysis of the simulated flows we performed a
decomposition into poloidal and toroidal components of the form

\begin{eqnarray}
\mathbf{u} &= \mathbf{u}_{tor} + \mathbf{u}_{pol} 
&= \nabla \times [\Psi(r,\theta,z) \mathbf{e}_z] + \nabla \times \nabla \times [\Phi(r,\theta,z) \mathbf{e}_z] 
\end{eqnarray}
with unique potential functions $\Psi$ and $\Phi$ and the unit vector
in $z$ direction $\mathbf{e}_z$.  To be precise, in the periodic box,
we normally need to add a component $F(z)$ depending only on $z$
(see \cite{sch1992} and (course 2, C. A Jones)
\cite{houchesdynamos}). Although our penalized cylinder is periodic
along the z axis there is no mean flow crossing the box as the gap
between the cylinder and the disks is small. We checked that $F$ is
zero (up to the numerical digit precision).

\begin{figure}
\includegraphics[height=4.5cm]{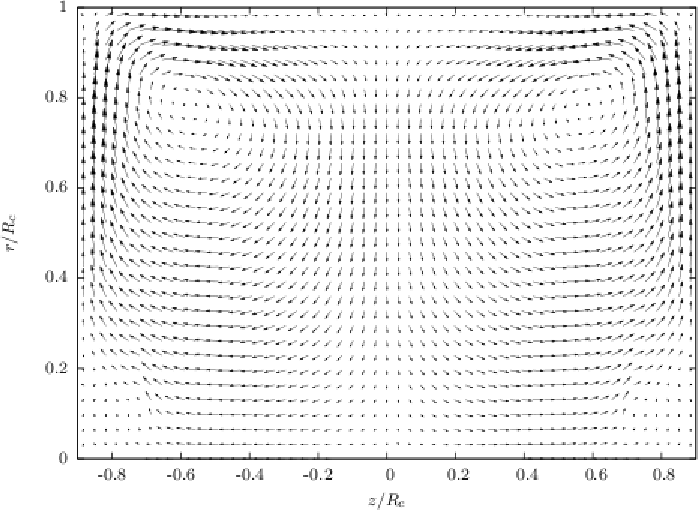}
\label{fig:torcyl}
\includegraphics[height=4.5cm]{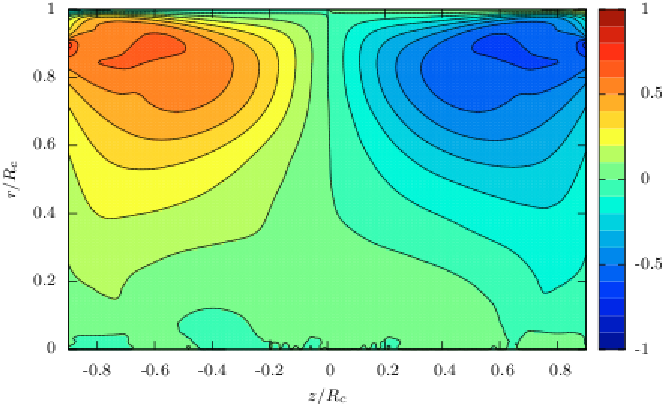}
\label{fig:polcyl}
\caption{Projections of the poloidal (left) and toroidal (right)
  components generated by impellers with curved blades and positive
  direction.}
\label{fig:pol-tor}
\end{figure}

\begin{figure}
\includegraphics[width=8cm]{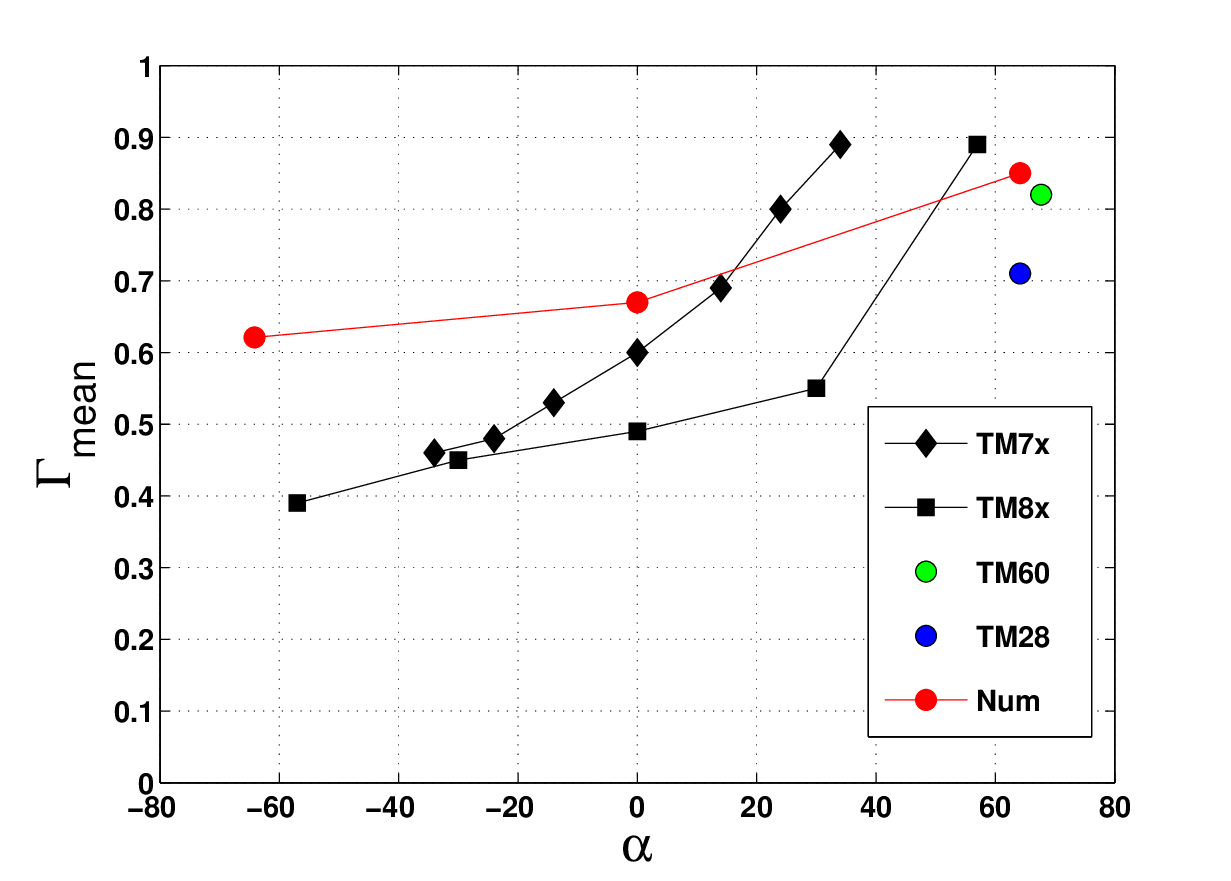}
\includegraphics[width=8cm]{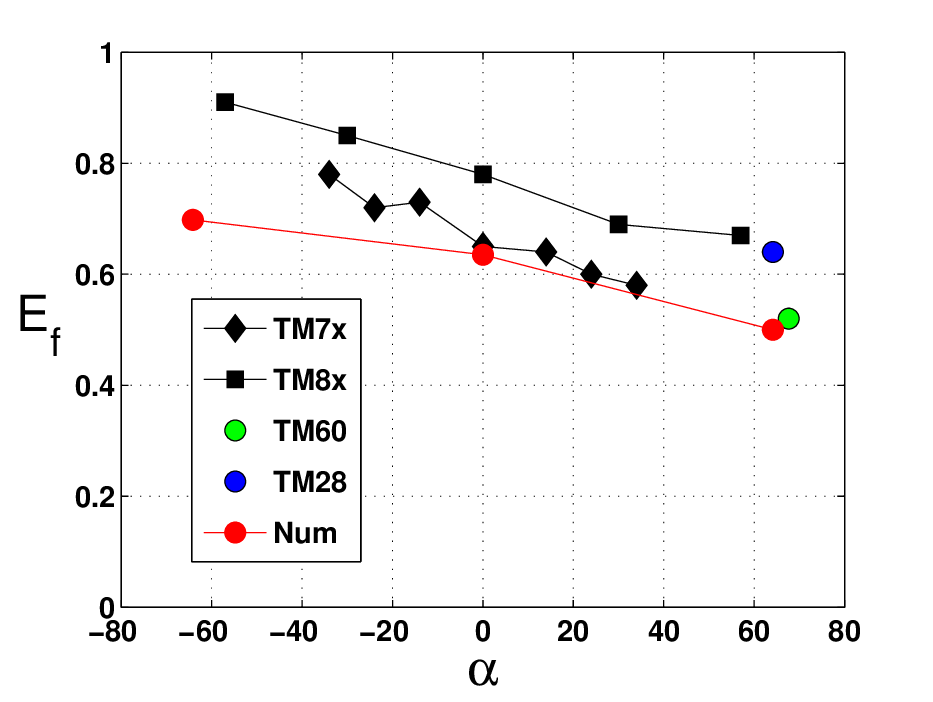}
\caption{
From different water experiment setups given in data tables (TM7x \cite{ravelet_toward_2005}, 
TM8x \cite{ravelet_these_2005}, TM60 and TM28 \cite{marie_numerical_2003}) 
and from our numerical results, we plot
(left) the ratio of the poloidal and the toroidal mean velocity ($\Gamma_{mean} $) 
and (right) the efficiency  $E_f= \nicefrac{U_{max}}{\Omega R_c}$  of the impellers
both versus the flow expulsion angle $\alpha$ at the end of the blades.
}
\label{fig:gamma-Ef}
\end{figure}

This decomposition has also been performed with the experimental data
with the TM28 impellers, which allows a comparison of experimental and
simulated data. Supposing axial symmetry around the z axis, we can
compute the poloidal and toroidal two
dimensional fields $\nabla \times [\Psi(r,z) \mathbf{e}_z]$ respectively
$\nabla \times \nabla \times [\Phi(r,z) \mathbf{e}_z]$ (see figure
\ref{fig:pol-tor}) and compare them to
figure 3 of \cite{marie_numerical_2003}.  We have visually a good
agreement, the poloidal flow consists of two large scale vortices in
the regions $z<0$ and $z>0$, where the toroidal components
respectively point in opposite directions.  Impellers with different
curvature (negative curvature and straight blades) produce the same
kind of flow structure.  From the images it is difficult to
distinguish between the different blades configurations.  We therefore
present in table~\ref{table:pol-tor} the mean and maximum velocity of
the poloidal/toroidal components, all of them rescaled by their
respective maximum velocity of the impellers $V_{max}=\Omega R_d$.
\begin{table} 
\begin{tabular}{|c||c|c||c|c|c|c|c|c| }
\hline
          &       \bf{TM28} &  (+)  &  (-)  &  \tiny{(straight)} & (+) 512 & (+) $1.6\Omega$ &  (+) $\nu/2$   \\
\hline
$u_{pol,mean}$ &    0.199     & 0.174 & 0.141 & 0.164     & 0.184   & 0.166          & 0.179 \\
$u_{pol,max}$  &    0.492     & 0.425 & 0.380 & 0.453     & 0.443   & 0.452          & 0.460 \\
$u_{tor,mean}$ &    0.281     & 0.205 & 0.228 & 0.245     & 0.217   & 0.202          & 0.217 \\
$u_{tor,max}$  &    0.691     & 0.535 & 0.824 & 0.708     & 0.538   & 0.530          & 0.509 \\ 
\hline
$\Gamma_{mean}$ &   0.71     & 0.850 & 0.621 & 0.670    & 0.847   & 0.822           & 0.825 \\
$\Gamma_{max}$  &   0.71     & 0.794 & 0.461 & 0.640    & 0.824   & 0.854           & 0.904 \\
\hline
$T$ & - & 1.000 & 1.295 & 1.375 & - & - & - \\
\hline
\end{tabular}
\caption{Quantities from experiments and simulations :  
the maximum and the mean of the poloidal and toroidal velocity and the respective ratio    
$\Gamma_{max} =\nicefrac{u_{pol,max}}{u_{tor,max}}$ and $ \Gamma_{mean} = \nicefrac{u_{pol,mean}}{u_{tor,mean}}$.
All the velocities are normalized by the maximum velocity of the impellers $V_{max}=\Omega R_d$.
A quantification of the poloidal and toroidal components is done by extracting the maximum 
and mean values in the bulk, the region $-0.8 R_c < z < 0.8 R_c$.
In addition, the average torque $T$ on the impellers has been computed 
for the three simulations with lower resolution and normalized
to the value obtained in the run (+).
}
\label{table:pol-tor}
\end{table} 
Those values can be compared to the experimental
ones~\cite{marie_numerical_2003}.  Our velocities of the $(+)$
configuration are $10\%-20\%$ lower than the experimental data of
``TM28''.  This could be explained by the fact that our efficiency
coefficient is lower than the ``TM28'' configuration (see figure
\ref{fig:gamma-Ef} right), implying a smaller velocity in the central
region of the vessel.  We also compare the ratio of the poloidal over
the toroidal velocity versus the expulsion angle of the blades
$\alpha$ with different water experiment results
\cite{ravelet_toward_2005,ravelet_these_2005,marie_numerical_2003}.
This ratio was used to seek the dynamo onset as a control
parameter. Like the efficiency the ratio $\Gamma_{mean}$ for the $(+)$
configuration is closer to the ``TM60'' than the ``TM28'' setup (figure
\ref{fig:gamma-Ef} left). However we stress that our ratios have also
a positive slope. Even if there is not a perfect agreement our ratios
are quite close to the different experimental measurements.

\subsection{Impact of viscosity and rotation speed}\label{sec:resolution}

In order to study impact of the viscosity and the disk rotation rate
on the mean flow structure we performed additional simulations with
curved impeller blades and positive turning direction.  In the first
of these runs we increase the resolution to $512^3$ grid points (Run
$(+)~512$ in table \ref{tab:datas}\&\ref{table:pol-tor}), while all
other parameters are unchanged.  This simulation can be seen as a
convergence test. The velocities for the toroidal and poloidal
component are slightly closer to the experimental values, but the non
dimensional quantities, especially the poloidal-toroidal ratios $\Gamma$, 
are unchanged, showing that at
$256^3$ grid points, our simulations are already converged.
 
In the second run (run $(+)~1.6\Omega$ in tables
\ref{tab:datas}\&\ref{table:pol-tor}), the angular velocity of the
impellers is increased by a factor of 1.6, while all other parameters
as well as the resolution are unchanged. The Reynolds number then
increases to $Re=3888$. The listed values remain constant, only the
ratio of the rotation period over the eddy turn over time slightly
decreases.

In the last run the viscosity is lowered by factor 2 (run $(+)~\nu/2$
in tables \ref{tab:datas}\&\ref{table:pol-tor}) and the resolution is
increased to $512^3$ grid points. In this case the Reynolds number
increases up to $Re=4860$. Regarding all quantities obtained in the
simulations, there is evidently no major influence of rotation speed
and viscosity on the mean flow profile.  Only the fluctuation rate
$\delta$ slightly increases.

The mean flow quantities which we present do not change with the
rotation rate or the viscosity. This implies that the corresponding
simulations are already in an asymptotic regime, where the Reynolds
number has only little effect on the large scale structure of
the flow. Indeed, the range of Reynolds number numerically acheived in our work ($2500-4800$), 
is at the edge the inertial regime of water experimentals results  
(\cite{ravelet_supercritical_2008} see their Fig. 5 and Fig. 7).
In this experimental campain, the Reynolds number has been increased progressivelly
by growing the rotation frequency of the disks,
to explore different phases : viscous, transition and inertial regimes.

\section{Local near-blade structures}\label{sec:local}

\subsection{Vortex generation behind blades}\label{sec:helicity}

Besides the general flow structure imposed by the impellers, our main
interest lies in the near blade flow in the frame of reference of the
rotating impeller. Here, we present and analyze the structures
obtained in the three different configurations (($+$), ($-$), straight)
with $n=256$. We construct the mean flow in the rotating frame of
reference by first averaging the flow each time the blades pass the
same positions, which is every eighth of a turning period. As we
expect to find the same mean flow at each blade we also take the
spatial average for rotations around $90^\circ$. As the running time
of our simulation corresponds to 20 impeller turns we were able to
average over 640 realizations. The mean flow in the rotating frame is
obtained from this average by subtracting a solid rotation
$\mathbf{u}^\ast = \mathbf{u}-\mathbf{\Omega}\times\mathbf{r}$.

\begin{figure}
\includegraphics[width=8cm]{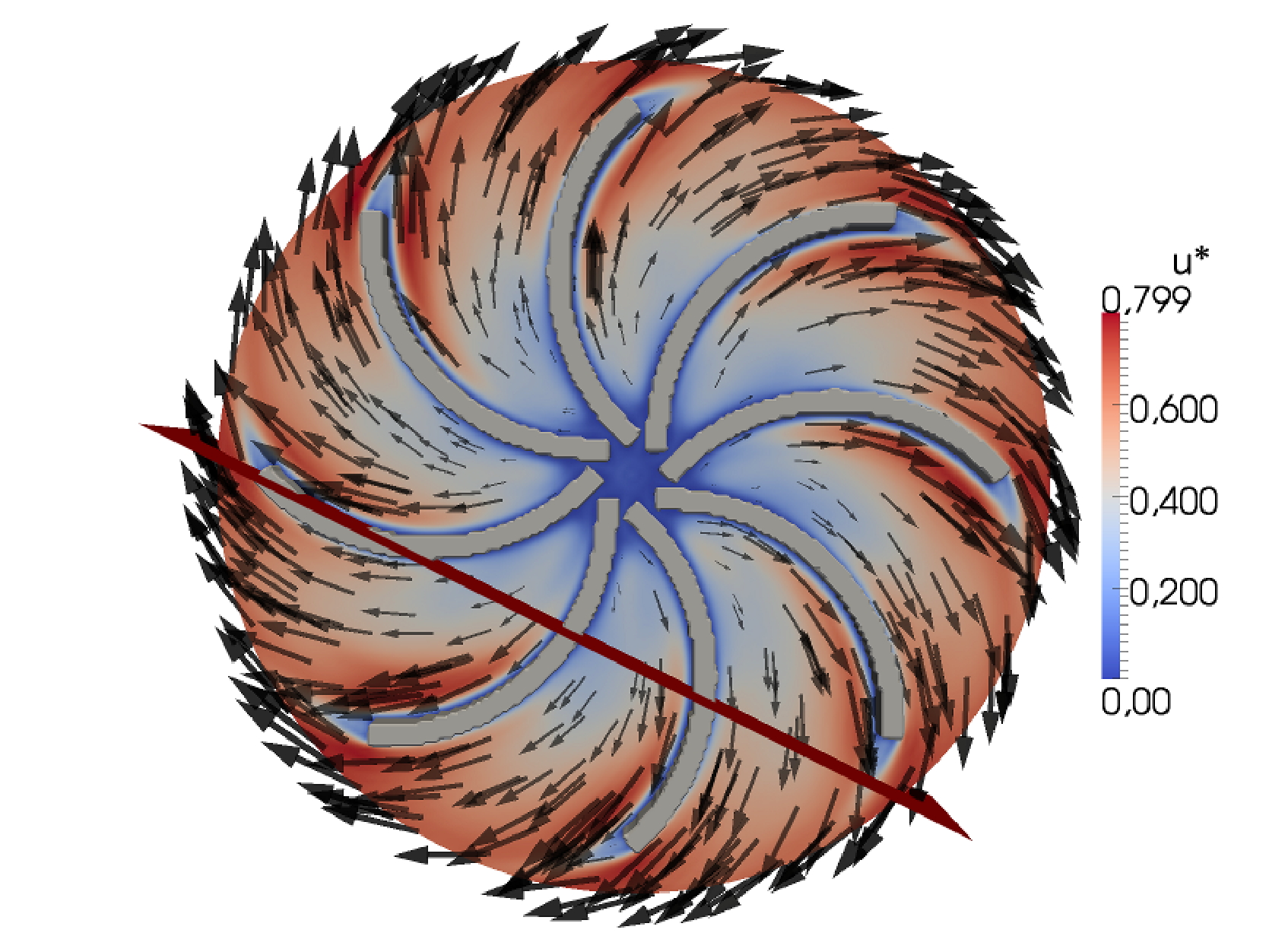}
\includegraphics[width=8cm]{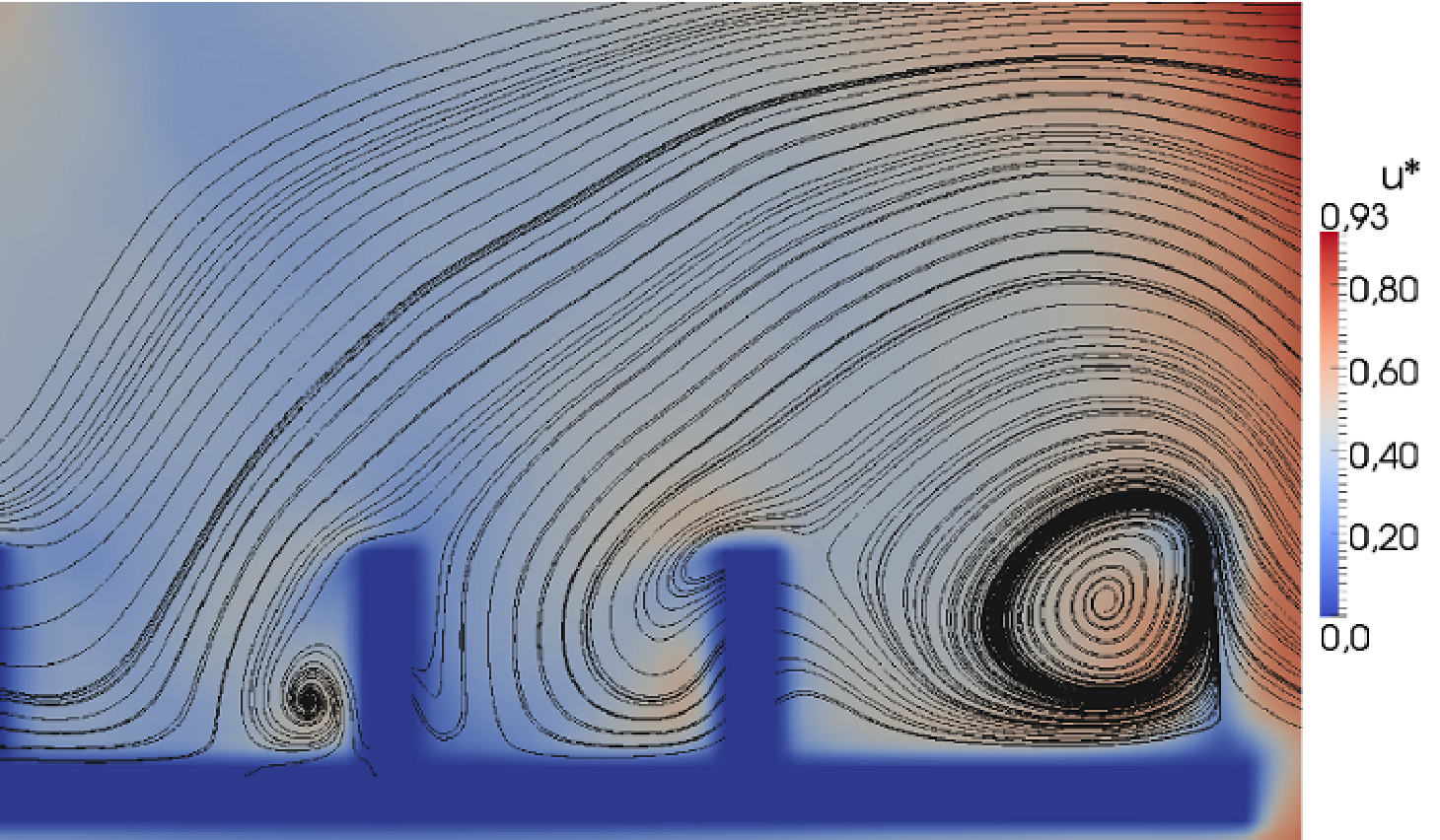}
\includegraphics[width=8cm]{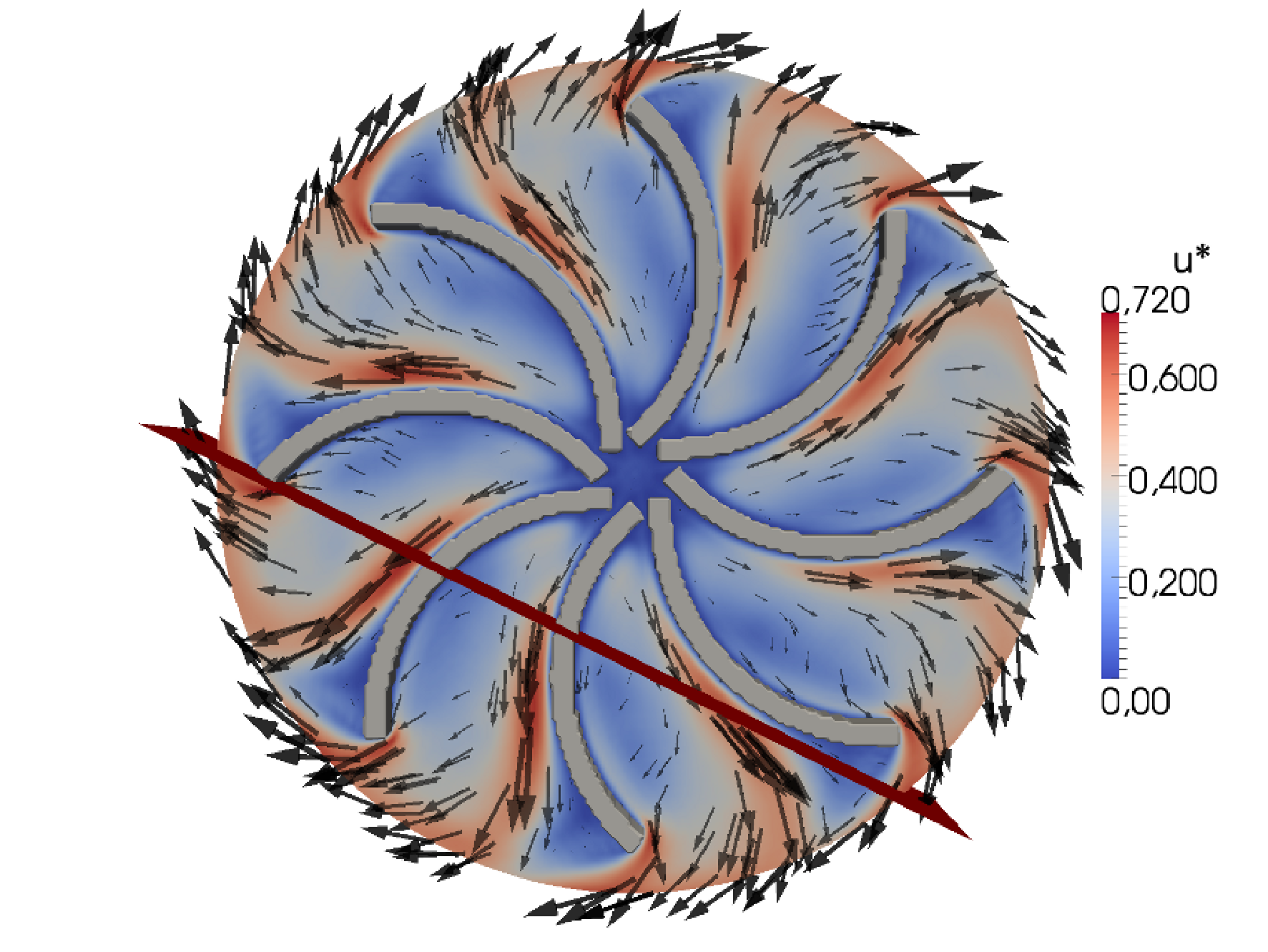}
\includegraphics[width=8cm]{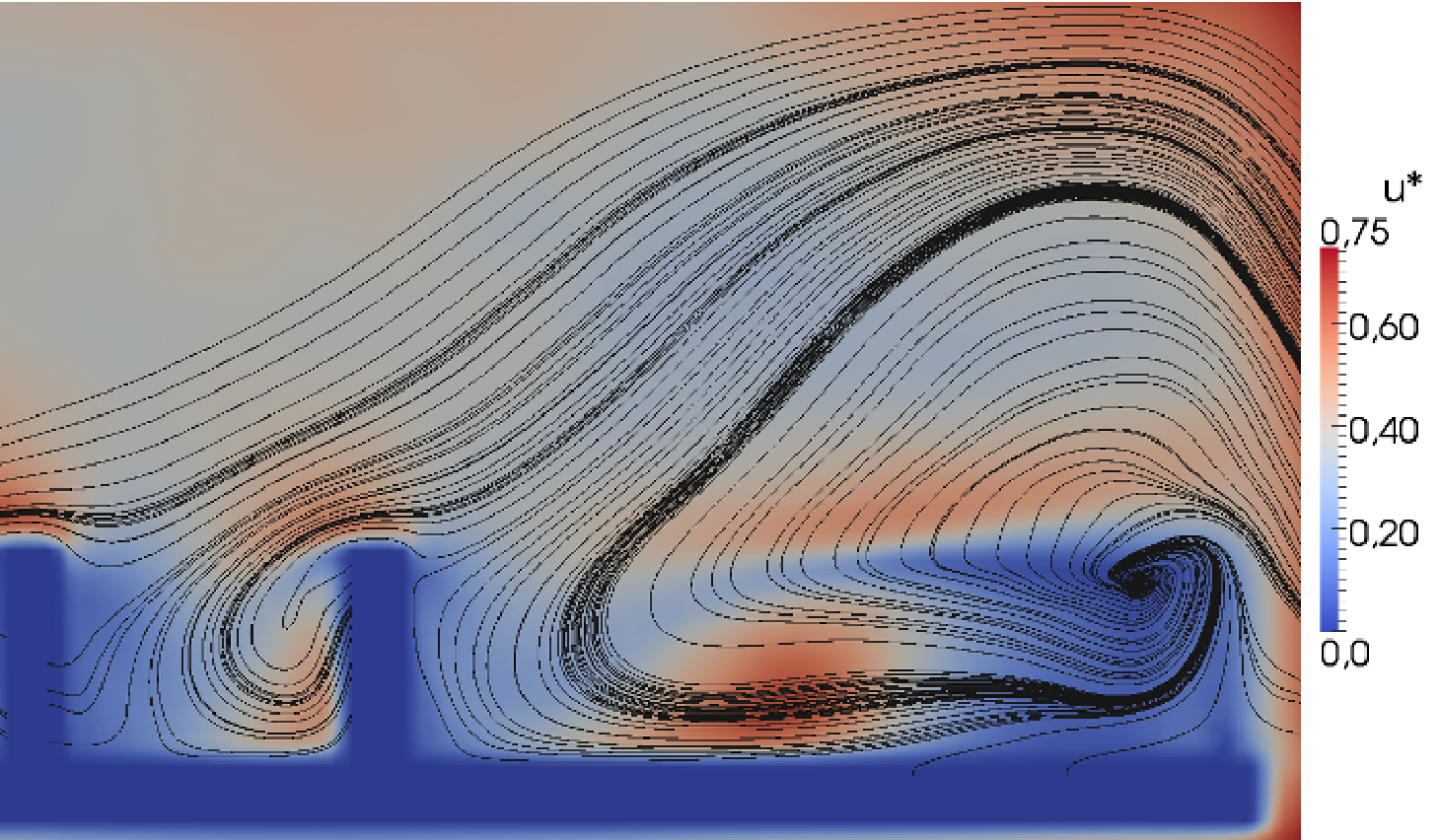}
\includegraphics[width=8cm]{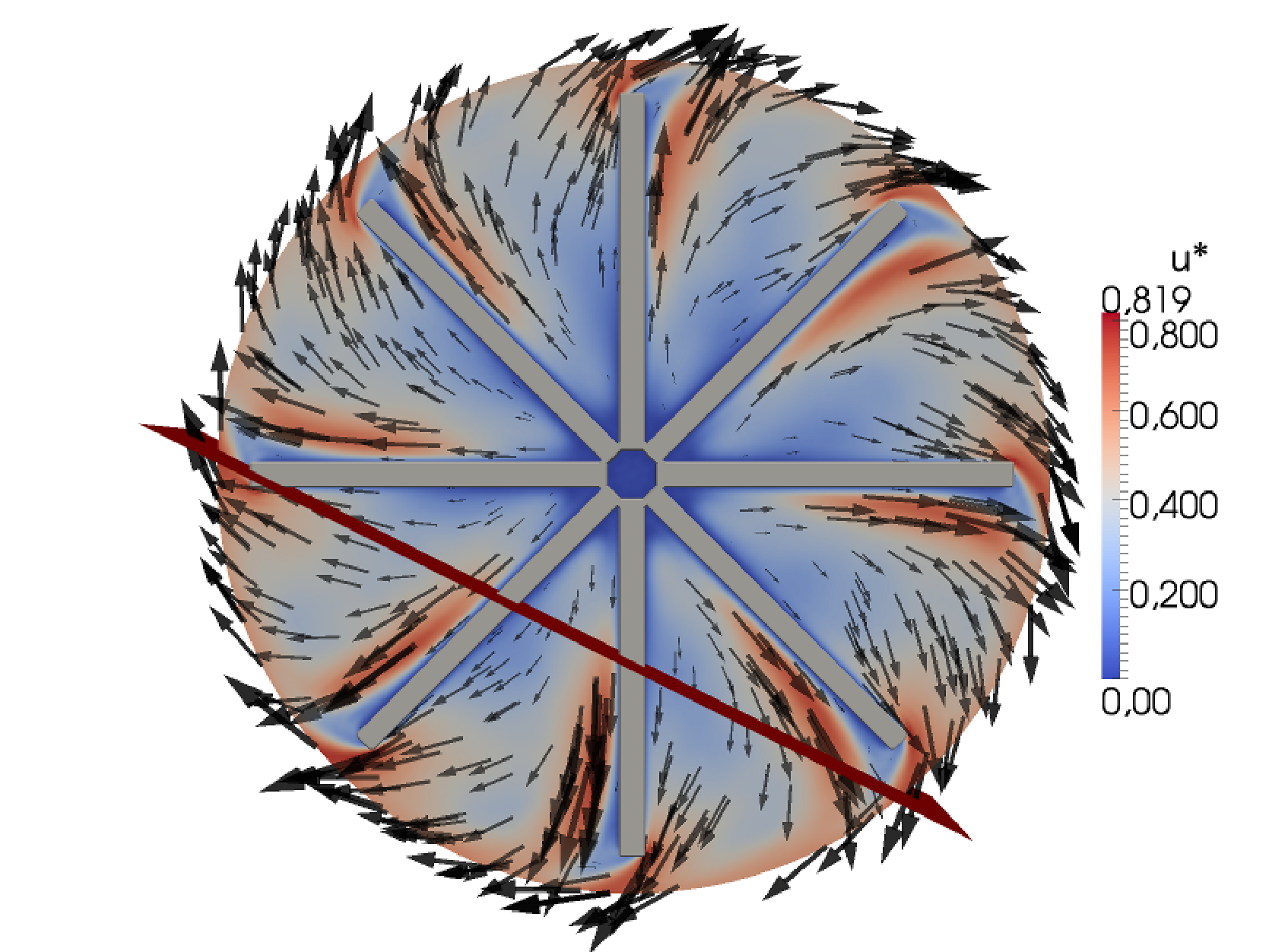}
\includegraphics[width=8cm]{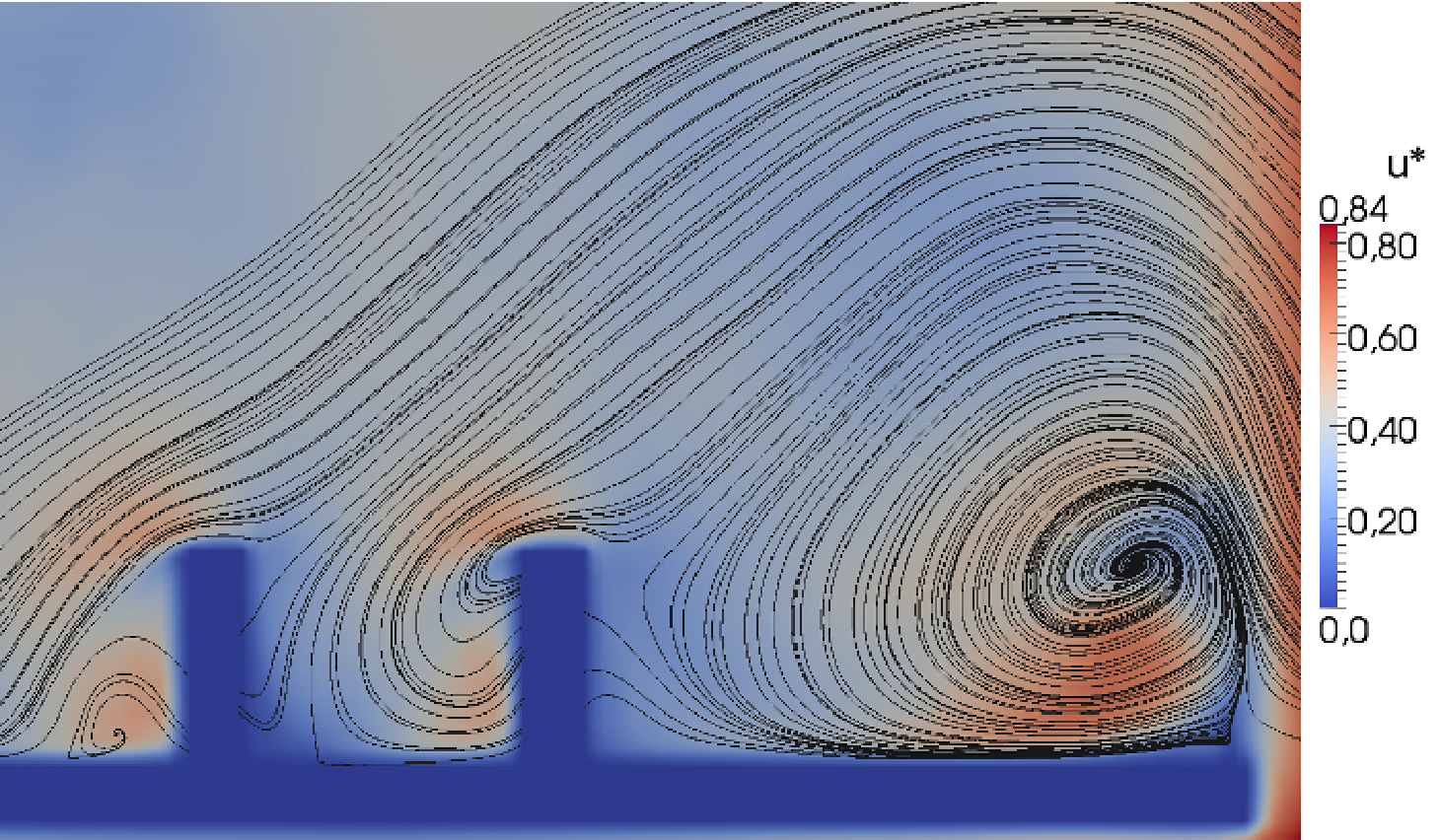}
\caption{\small Visualization of the velocity field on a top view (left) and
  the streamlines of velocity on a transverse view (right), for our
  three configurations (from top to bottom (+), (-), (straight)).  The
  velocity is averaged in time, projected on the plane, and computed in
  the rotating frame ($\mathbf{u}^\ast =
  \mathbf{u}-\mathbf{\Omega}\times\mathbf{r}$).  In all images, the
  color map represents the magnitude of the velocity.  The transverse
  view planes are perpendicular to the red line shown on the top view
  planes. Not all of the perpendicular planes associated with the red lines
  is plotted (only 70\% of the red line of the left side).  The
  position of the blades helps to relate the top and side views.
  Note that this projection of the velocity in the
  considered plane is not solenoidal, thus streamlines could end
  at the boundary, where the projected velocity tends to zero.
  Streamlines of the 3D flow do not enter in the solid object, 
  but instead slide along the boundary. 
  This behaviour is not captured by the 2D projection.}
\label{fig:flow_blade}
\end{figure}

To get clear pictures of the flow we consider two different planes:
one parallel to the disk plane, with a view from the top and the other
perpendicular to the disk (the position of the perpendicular plane is
indicated by a red line in the top view (left column of
figure \ref{fig:flow_blade})). In the top view, we show the velocity
projected onto the plane while its amplitude is represented by a
color map. This perspective shows how the flow is sucked in from above
to the center, moving between the blades and how it is finally
expelled from the disk outwards. Just by looking at the left column of
figure \ref{fig:flow_blade}, we can easily distinguish between the
different configurations, specially the $(+)$ and the $(-)$, where the
expelled flow in the rotating frame has the same rotational direction
as the disk rotation.  The horizontal component represents the major
part of the magnitude of the full velocity along the blade.  There is
clearly an acceleration from the center to the expulsion area. The
most rapid velocity is generally along the pushing blade wall. A
small sucking area is located just behind at the end of the blade.

In the perpendicular plane we decide to visualize the velocity by
streamlines to highlight the topology of perpendicular flows. Note
that the projected velocity is not solenoidal, thus streamlines may
have an end point. In all
three simulations, (right column of figure \ref{fig:flow_blade}) the
streamline plots show vortex rolls emerging directly behind the moving
blade, as vortices are ripped off at the blade's edge. Those vortices
appear to take most of the space between the blades.  Clearly, the
negative configuration (-) has a different topology than the two
others. This observation agrees with the horizontal velocity in the
top views. The cut along the red line allows the visualization of the
mean flow vortex at different radii for different cells. In the
positive configuration (+), it can be deduced that a cone vortex is
produced along the pushing blade in each inter-blade cell.  Those
vortices are also present in the straight configuration which is,
however, less clear for the negative configuration $(-)$.

\section{Discussions and perspectives}\label{sec:conclusion}

\subsection{Numerical and experimental comparisons}

By means of direct numerical simulations using a penalization
technique we are able to reproduce the large scale structure of
experimental von K\'arm\'an flows produced by moving impellers. Our
good agreement with the experimental results could be explained by the
fact that the mean flow geometry and other global quantities are
converging rapidly even at low Reynolds number.
  
The natural next step could be to seek the small scale properties of
the von K\'arm\'an flow, like the studies on filaments
\cite{douady_direct_1991,cadot_characterization_1995}, effect of
large scales on small scales \cite{labbe_study_1996} or the energy
injection \cite{mordant_characterization_1997,labbe_power_1996}.  Of
course, direct numerical studies with confined flows in the full von
K\'arm\'an geometry remain a challenge asking for a big increase of
spatial resolution. Our modelization using the penalization could be easily used to explore
different experimental setups - at least for the large scale
properties.

\subsection{Vortex and Dynamos}

We found a characteristic outwards spiraling vortex between the
blades.  Note that the influence of the vortex generated around the
blades is suspected to have a strong impact on the dynamo effect
\cite{petrelis_magnetic_2007}. Some numerical results assuming a
dynamo mechanism concentrated around the disk-blade structure
\cite{laguerre_impact_2008, giesecke_role_2010} have found that the
magnetic mode has a dipole structure according to the experimental results.
The geometry of the experimental magnetic mode cannot be explained by
a mean flow dynamo only. Recent numerical studies 
using FLUENT ($k-\epsilon$)-RANS simulations \cite{ravelet_kinematic_2012}
computed the $\alpha$-tensor produced by the vortex dynamics, 
given a switch between $\alpha^2$ and $\alpha-\Omega$ dynamo types.  

Despite these vortex dynamics, 
without soft iron impellers the dynamo threshold was not achieved.
The material of the impellers plays a crucial role for the efficiency of the dynamo
action~\cite{verhille2010,giesecke_influence_2012,miralles_dynamo_2013}.   
The material properties of sodium make in-situ diagnostics very difficult. Direct
numerical simulations provide a unique tool to assess spatially and
temporally resolved variables. With our dynamics of the blade vortex
and correct treatment of the magnetic properties of the solid impellers,
all the physical ingredients should be present to address those different questions, 
to seek the interaction between the conducting fluid and the ferromagnetic impellers. 
We could also check the different dynamo onset prediction or measurement of the different configurations 
produce by the variation of the blade geometries or material properties~\cite{miralles_dynamo_2013}.

\subsection{Long term dynamics} 

When the water experiments are running during a long time, the
von K\'arm\'an mean flow can change to different topology solutions,
breaking symmetries
\cite{ravelet_multistability_2004,ravelet_supercritical_2008,de_la_torre_slow_2007}.
In the experiment, the typical time scale to record this
multi-stability is around $10^5$ to $10^6$ hydrodynamic large eddy
turnover times.  In our present simulations, we are completely out of
reach to record such dynamics: we computed $20$ disk turns, which
represent around $45$ eddy turnover times only. It will thus be very
interesting, even if less popular than the highest resolution limit,
to perform long numerical runs with moderate grid points to reach and
study the long time physical behaviors or to improve statistical data.

\section{Acknowledgment}

We acknowledge fruitful discussions with Nicolas Plihon, Arnaud Chiffaudel.
Parts of this research were supported by Research Unit FOR 1048, 
project B2, and the French Agence Nationale de la Recherche 
under grant ANR-11-BLAN-045, projet SiCoMHD. 
Access to the IBM BlueGene/P computer JUGENE at the FZ
J\"ulich was made available through the project HBO36.
Computer time was also provided by GENCI in the IDRIS facilities and the Mesocentre SIGAMM machine,
hosted by the Observatoire de la C\^ote d'Azur.

\vspace{1cm}

\bibliographystyle{unsrt}
\bibliography{references}

\end{document}